\documentclass[a4paper]{article}
\usepackage[T1]{fontenc}
\usepackage[utf8]{inputenc}
\usepackage[italian,english]{babel}
\usepackage[autostyle,italian=guillemets]{csquotes}
\usepackage{multicol}
\usepackage{microtype}
\usepackage{guit}
\usepackage{booktabs}
\usepackage{graphicx}
\usepackage{mathptmx}
\usepackage{amsfonts}
\usepackage{amssymb}
\usepackage{amsmath}
\usepackage{ifpdf}
\usepackage{dcolumn}
\usepackage{textcomp}
\usepackage{tabularx,array}
\usepackage{ulem}
\usepackage{placeins}
\usepackage{epstopdf}
\usepackage{etex}
\title{Photometry Study of  Open Cluster NGC 7788} 
\author{Tasselli, D. \\ TS Corporation Srl - Department of Astronomy and Astrophysics \\ Regione Salamia, 10010 Andrate TO - Italy \\ E-mail:diego.tasselli@tscorporation.org} 

\begin{document}

\maketitle

\begin{center}

\end{center}
\begin{abstract}
A new CCD photometry  survey of the open cluster NGC7788 in U, B, V, R and I colors to magnitude V=20 is present. Extensive comparison of our photometry with other published dataset shows excellent agreement, indicating that CCD photometry is capable of producing a uniform set of measurements consistent with the photometric system defined primarily by the Landolt standard sequence. For this study we used the 113 stars considered to be cluster members by Becker on 1965. Assuming R = 3,06 at (B-V) = 0.0 and the Zero Age Main Sequence of the Pleiades as reference locus with distance modulus 5.57 (Walker 1985), we have derived from the resulting color-magnitude diagrams new and more accurate estimates of the reddening and distance modulus of the cluster with turn out to be E(B-V) = 0,28$\pm 0,03 $ and $ (m-M)_{o}$ = 11,9 $\pm$ 0,24 respectively. With the parameters obtained and assuming solar compositions, the age of NGC7788 is $9,33\times10^{7}$ yr, and the distance is 2398 parsec.
\end{abstract}
\begin{multicols} 
{2}

\section{Introduction}
Open cluster are good tools to analyze the large-scale proprieties of the disk of our own galaxy and the test the theories of stellar and galactic evolution. In this first paper we present the result concerning NGC7788, a cluster located: R.A.:$23^h56^m.4$, \ Decl..$+61^\circ 23^m.5$ near the Perseus spiral arm. Our objective was to produce a deep U,B,V,R,I survey of a region centered on the asterism of NGC7788, calibrated measurements according to the standard U,B,V,R,I Landolt, using the earlier study of NGC7788 done by Becker, associated with equatorial coordinates and cross-references to the mentioned investigation, as Landolt stars \cite{Landolt:ref1}.. This new photometry would be immediately applicable for comparison with theorical isochrones, and for the investigation spatial distribution, and luminosity function of the cluster. The scope of our survey was limited primarily by the size our telescope and the observing time available. The resulting photometry should, in short, give us a fresh look at a cluster. 
\section{The Data} 
\subsection{Observations} The cluster was observed in 2013 February (UT) with the Richey-Chretien telescope of the TS Corporation on Andrate (TO) - Italy station, equipped with a CCD camera (FLI EEV2 back illuminated,  2048$ \times $2048  pixel $ \mu$m $0,39$ arcsex/pix) and set of UBVRI filters (Johnson-Kron-Cousins). Except for the last hour of the final night, when high cirrus clouds moved in, the weather was exceptionally clear. Fig. 1 show the identification map for the stars measured and the Table 1 gives the correspondence between our star numbers and result of analysis data. Table 2 show the journal book of observations and capture image. The caratteristic of strument are visible on this paper:\cite{Tasselli:2011ug}. Preliminary processing of all CCD frames, to apply bias and flat field corrections, was alone with standard routines in the IRAF software package.  For each evening, at least 28 flat field exposures were available in each filter U,B,V,R and I, from dome illumination, and at least three frames each in U,B and V using sky illumination. The dome flats were applied to all image but for the U,B, and V images the dome flats  lefts slight residual patterns due presumably to the illumination or color balance problems. The IRAF sky illumination or color balance problems. No residual patterns can be seen on the images. The photometric reduction were made with a software package, IRIS, developed at Christian Built. The IRIS is similar to several other photometry programs in using a point-spread-function (PSF) fitting procedure to calculate the magnitudes of all stars on a CCD image. Once a model PSF is computed, based on the observed image profiles of the brighter, relatively isolated stars on the frames, a map of the cross-correlation function between the PSF and the original image is used to identify all measurable stars on the frame. The magnitudes of stars in the Table 1 are computed by fitting the position and scale of the PSF to each star image in turn, in order of decreasing brightness. The Zero Point of the frame is set during the PSF calculation, thought aperture photometry of the stars used to calculate the PSF. 
\begin{tabular}{l|c|r|r}
\hline
Date &Filter &Time &Airmass \cr
\hline
feb-12 &U &$120 \sec$ &1,472166 \\
&B &$30 \sec$ &1,52828 \\
&V &$30 \sec$ &1,556142 \\
&R &$10 \sec$ &1,563522 \\
&I &$20 \sec$ &1,607856 \\
\hline
\end{tabular}
\subsection{\normalsize{Transformations and Reductions}}
Instrumental magnitudes for all measured stars were transformed to a standard system using fitting coefficients derived from observations of standard stars whose magnitudes have been well established in earlier studies. Our primary source of standards compiled by Landolt (1983). Transformation coefficients between instrumental and standard magnitude was determined using the following equations: 
\begin{equation}U=M_{u}+ U_{o}+y^U_{U-B}(U-B)\end{equation}
\begin{equation}B=M_{b}+ B_{o}+y^B_{B-V}(B-V)\end{equation}
\begin{equation}V=M_{v}+ V_{o}+y^V_{B-V}(B-V)\end{equation}
\begin{equation}R=M_{r}+ R_{o}+y^R_{V-R}(V-R)\end{equation}
\begin{equation}I=M_{i}+ I_{o}+y^I_{V-I}(V-I)\end{equation} 

\noindent M$_{u}$, M$_{b}$, M$_{v}$, M$_{r}$, M$_{i}$ are the instrumental magnitudes; U,\,B,\,V,\,R,\,I are the standard magnitudes, U$_{o}$, B$_{o}$, V$_{o}$, R$_{o}$, I$_{o}$ are the Zero Age Main calculated with the Pleiades point; $y^A_{A-B}$ are the Index of colour determined by calibrating the standard deviations of the instrumental magnitudes with the standard stars. To determine the zero point and the color index relative to each equation band present in the article, we used the method of least squares. This allows us to represent the distribution of stars along a straight line in which the zero point is the intercept term and the color is the slope. Taking labeled equation number 1-5 we have: \\
$ x_{i}$ = color band and $y_{i}$ = photometric band 
$\Delta= \bigl[N \ \sum x_i^2-\Bigl[\sum x_i\Bigr]^2$
\\
and Index of Color from: 
\\
$y^A_{A-B}=\Bigl[N\sum x_iy_i-\sum x_i\sum y_i\Bigr] / \Delta$ 

\noindent The index of colour used for transforming to the U,B,V,R,I standard system for the observing run were 0.138 in V, 0.292 in B, 0.769 in U, -0.519 in R, and 0.442 in I. Fig.2(a)-(e) show the index of colour. The standard deviations of the magnitudes of the stars used in transforming to the U,B,V,R,I system were 0.213 in V, 0.230 in B and 0.196 in U, 0.330 in R, and 0.329 in I. Fig. 3(a)-(e) show the standard deviation. For the R and I filter, we used only the instrumental values because not having available these values in the work of Becker. Instrumental magnitudes for all stars on our image were transformed to the U,B,V,R,I system by using Eqs(1)-(5). The photometry for all stars is show in Table1: Column 1 give the stars numbers. Column 2 and 3 give the X center and Y center Coordinate about this study in V magnitude. Column 4,5 and 6, give the V, B-V and U-B data photometry measured of this study. The last column give the cross reference stars with Becker study.
\subsection{Photometric Error} The precise definition of photometric errors in the measured of star image on CCD-frames is not an easy task. In fact, the ''total'' error is produced by the contamination of at least three different contributions:
\begin{itemize}
\item the measuring error in each individual frame and in the standard stars used for the calibration (external error)
\item the intrinsic error involved in the Standard System
\end{itemize} 
Since repeated fits of any star on the same frame always give an identical numerical solution, the only available determination of the internal errors is represented by the plots of the colour correction for the standard stars used for the calibration.
\subsection{Comparison with Previous Studies} Despite of its position, very close to the Standard cluster NGC 7790, NGC 7788 seems at this moment, very little studied. The only previous photometric study of this cluster based, between other things, only on photographic photometry, is that of Becker (1965). Such photometry has been publisched in \cite{Becker:ref1}. We are confident that the use of a CCD detector should guarantee the measures presented in Table 1 even at the faint end. In 1965 Becker obtained for this object the following two important data, a reddening value of 0.28 and a cluster distance of 2410 parsec. Table 3 show the final values of mean colour excesses and the true distance moduli, along with their rms deviations. 
\section{Data Analysis} 
\subsection{\normalsize The coulor-magnitude diagrams} Fig. 4, 5, 6, 7, 8 and 9 show the (V, B-V), (V, V-R), (R, R-I), (U,U-B), (V, V-I) and (V, U-B) diagrams of NGC7788 resulting from the described observations and reductions of this study. Fig10 show a comparative with this study and Becker study for this open cluster. 
\subsection{Field stars} NGC7788 is near the Perseus arm. This location contributes to give a sizable contamination of foreground and background objects. Since no reliable radial velocities are available (except for the variable stars), the discrimination of non-members from the observed sample can hardly be carried out by checking each star individually on purely photometric grounds. In other to get anyway a sample less contaminated than that directly obtained from the reductions, we have thus decided as a first step to determinated the blue envelope of the Main Sequence (MS) in two Color magnitude Diagrams (CMDs) and to preliminary label as field stars those redder that the colour of the obtained by $\delta (B-V)> 0.3$ and $\delta (V-I)> 0.3$. This chose should not reject cluster members (at least for stars well below the turnoff) and yields a sample suitable for further analysis.
\subsection{Candidate members} With the cautions discussed in the previous sections, we can define a Main Sequence (MS)-band which includes the "candidate" members. The distribution of stars across the Main Sequence (MS) in the V us (V-R) color magnitude diagram was determined by applying a best fitting procedure with the Pleiades and Hyades ZAMS in (V, B-V) and (V, V-R) planes. Fig. 5 show the (V, V-R) diagrams. It's includes only stars of V magnitudes.
\section{Cluster Parameter} \subsection{Reddening and \\ Distance modulus} Open cluster are very frequently affect by large colour excesses, and variations of extinction across the cluster itself are often present.  In order to fit the observed Main Sequence (MS), determinated from Colour Magnitude Diagrams (CMDs) of the cluster described in the previous sections and to derive the reddening and distance modulus of NGC7788, we adopted the Pleiades sequence derived by Turner \cite{Turner:ref1} \cite{Turner:ref2} (1976,1979) as fiducially ZAMS rather than the Hyades sequence, to avoid the problems of controversial metallicity and age corrections. The adopted distance modulus of the Pleiades is 5.57$\pm$0.08 \,(Feast\& Walker 1987) \cite{Walker:ref2}. For the radio of total to selective absorption we have adopter the expression (Olson 1985 \cite{Olson:ref1},Walker 1987b) \cite{Walker:ref3}: \noindent $R=A_{V}/E(B-V)=3.06+0.25\,(B-V)_0\,+\,0.05\,E(B-V)$ \\
for the relation between the B-V and V-I colour excesses (Walker 1985) \cite{Walker:ref1}: \newline
\noindent $E(V-I)=1.25\, E(B-V)\,[1+0.06\,(B-V)_0+\,0.014E(B-V)]$ \\
\noindent We have then assumed for the mean intrinsic colour of the cluster $(B-V)_o=0.0$ which implies R\,=\,3.06 for E(B-V)\,=\,0.0. These choices allow the determination of both reddening and distance modulus through the two available plots, namely (V, B-V) and (V, V-I), where the best fitting procedure can be carried out. Of course, since the two plots are not totally independent and, moreover, we do not know precisely Y and Z, the solution is not unique. With these caveats, we find that the combination of values leading to the best fit to both the (V,B-V) and (V, V-I) sequences is $E(B-V)$=0.28$\pm0.03$ and a distance modulus $(m-M)$=11,9$\pm0,24$, where the error in the distance modulus results from the combination of the error in R, in E(B-V), and in the ZAMS fit. 
\subsection{The Luminosity Function} Determining the number of stars as a function of luminosity in a cluster is a difficult procedure because of the presence of background stars in the same line of sight. In his photographic survey, Racine (1971) used a comparison field outside the cluster to determine the amount of field star contamination. Francic (1989) used proper motion studies to determined membership. Neither method was suitable for this study, since our field did not go out even to the cluster edge, and since the available proper motion data does not go faint enough. Instead, we applies a statistical correction to our data based on the distribution of star in the colour-magnitude diagram(CMD). The method involved counting the number of field stars of a given magnitude with a region just blue of the main sequence(MS). The resulting brightness distribution of filed stars was then used to correct the raw data to determine the number of main sequence(MS)stars with each magnitude bin. The agreement with published luminosity function is quite good, and a comparison of our luminosity functions is show in Fig 11. \subsection{Age Determination} To determinate the age of NGC7788 we have applied an expansion of the classical method of isochrones fitting assuming as know data Y, Z, E(B-V) and distance modulus. Our method is based on the comparison of the observed (CMDs) Colour Magnitude Diagram of the cluster with the theorical diagram resulting of an object containing the same number of stars above the limiting magnitude as observed in actual cluster. Our simulated diagrams are based on two homogeneous set of stellar evolution models with solar metallicity $Z=0.02$, mixing length parameter $\alpha=1,5$ and intermediate mass loss $\eta=1/3$.  The first set of models takes overshooting from convective cores into account and range from $1,2 $ to $ 100\odot$ the others are standard models without overshooting and range from $0,7 $ to $ 120M\odot$ The simulated diagrams has been transformed into the (V, B-V) plane by applying to each star Walker's (1987b) relation for the total to selective absorption and the values of reddening and distance modulus of NGC7788 found in the previous section. We have used for the comparison only the 113 stars considered to be as actual members of NGC7788. Our best simulated distributions of NGC7788 based on standard evolutionary tracks corresponds to an age of $9,33x10^7$ years, significantly lower than the $1,58x10^8$ yr derived it overshooting is taken into account, and a Log (Age) of 7,97.  
\section{Conclusion}
This first CCD photometry of NGC7788 has allowed us to put more constraints on the main parameters of the cluster. By carrying out main Sequence (MS) fitting in the (V, B-V) and (V, V-I) planes\, we have determined the reddening and the distance modulus of this cluster. From a mosaic of CCD Image of NGC7788 cluster we have derived the following information:
\begin{enumerate}
\item We have tabulated U, B, V, R, and I colour of 113 stars. The photometry is presented with equatorial coordinates for each stars and cross reference to other studies to make it handy and useful for further research on the cluster. A comparison with previous studies shows that the photometry fits well into the standard system define by Landolt.
\item We present a sample of stars suitable for use a standard calibration for general CCD photometry.
\item Based on colour-colour plots of the cluster,\,we derived a reddening,\,E(B-V) of $0,28\pm 0,03$, a metallicity, [Fe/H]=-2, a distance modulus of 2398 parsec and a Log(Age) of 7,97.
\item Finally, the age of NGC7788 has been determined by means of simulated Colours magnitude Diagrams (CMDs) based on stellar evolution models either with overshooting from convective cores age $1x10^8$ yr or without $9,33x10^7$ yr.
\end{enumerate}
Give the importance of this cluster and the still insufficient set of data available, further observations on metal content and membership are needed.

\end{multicols}
\newpage
\begin{figure}
\begin{center}
\includegraphics[width=0.7\textwidth]{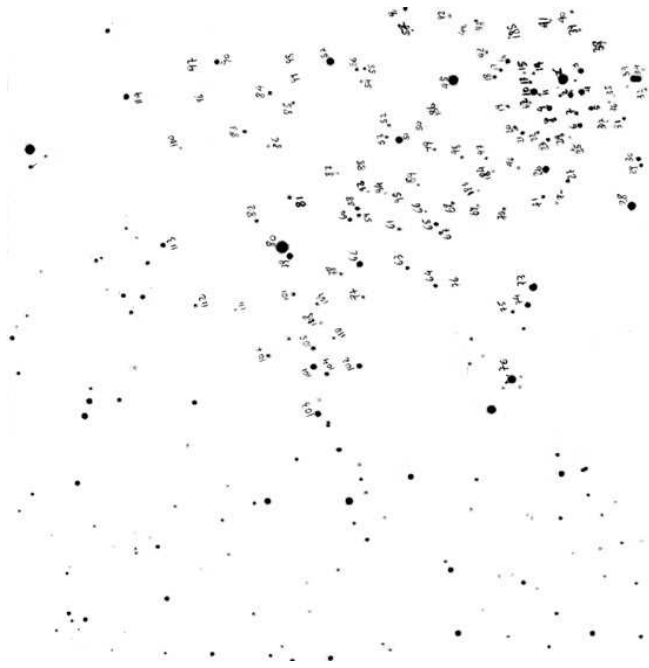}
\caption{Map of the stars measured in our filed of NGC7788. The identification numbers are from Table 1. Orientation: north is at the bottom, east on the right.}
\end{center}
\newpage
\end{figure}

\begin{table}
\caption{Table1: Data on CCD photometry in NGC7788} 
\tiny
\begin{tabular}{|r|r|r|c|c|c|c|c|c|c|c|c|c|r|}
\hline
\bf ID& \bf X Center& \bf Y Center& \bf Mv & \bf Mb & \bf Mu & \bf Mr & \bf Mi & \bf B-V & \bf U-B & \bf V-I & \bf R-I &  \bf V-R &\bf Ref \\ 
\hline
\bf 1 &1.189.706 &24.442 &9,51 &9,70 &9,12 &11,69 &9,50 &0,18 &-0,58 &0,01 &2,19 &-2,18 &1 \\ 
\bf 2 &559.710 &12.768   &12,70 &12,83 &12,78 &10,48 &11,73 &0,14 &-0,05 &0,96 &-1,26 &2,22 & 6  \\ 
\bf 3 &1.195.000 &31.446 &12,64 &12,81 &12,70 &14,25 &10,50 &0,17 &-0,11 &2,14 &3,75 & -1,61 & 2 \\ 
\bf 4 &250.380 &20.641   &15,21 &16,98 &- &10,25 &14,02 & 1,77 & - & 1,19 & -3,78 & 4,96 & 4 \\ 
\bf 5 &53.507 &17.672    &15,66 &16,82 &- &11,01 &10,31 & 1,17 & - & 5,35 & 0,69 &  4,65 & 3 \\ 
\bf 6 &1.029.269 &21.550 &15,51 &16,27 &- &10,68 &10,79 & 0,76 & - & 4,72 &-0,11 & 4,83 &  18 \\ 
\bf 7 &900.029 &17.502   &15,57 &16,32 &- &12,24 &10,33 & 0,75 & - & 5,24 & 1,91 &3,33 & 19 \\  
\bf 8 &249.879 &18.743   & 14,31 &15,19 &15,36 &10,03 & 11,91 &  0,88 &0,17 & 2,41 &  -1,87 &4,28 &  44 \\ 
\bf 9 &162.516 &27.072   & 12,88 &13,08 &13,05 &14,30 &10,01 & 0,20 &-0,03 & 2,87 & 4,29 &  -1,41 &  43 \\ 
\bf 10 &933.405 &36.887  & 11,64 & 11,69 &  11,17 & 12,34 & 12,78 &  0,06 &-0,53 & -1,14 & -0,44 & -0,70 & 10  \\ 
\bf 11 &682.950 &40.198  &12,36 &12,52 & 11,98 & 14,51 &10,28 &0,16 &-0,54 & 2,08 &4,23 &-2,15 & 5 \\ 
\bf 12 &979.390 &52.059  &14,57 &14,86 & 15,09 &10,89 &14,20 &0,29 &0,23 & 0,37 &-3,32 &3,69 &36 \\ 
\bf 13 &1.111.380 &53.185 &14,57 &15,00 &15,24 &10,22 &11,04 &0,43 &0,25 &3,53 & -0,81 &4,35 &11 \\ 
\bf 14 &246.317 &56.017  &15,22 &15,86 &- &11,28 &11,08 &0,65 &- & 4,14 & 0,20 &3,94 &13 \\ 
\bf 15 &329.621 &69.044  &15,84 &16,89 & - & 10,90 &10,85 &1,05 &- & 4,98 &0,04 &4,94 &14 \\ 
\bf 16 &1.146.342 &82.014 &13,61 &13,85 &13,70 &10,28 &10,22 &0,24 & -0,15 &3,39 & 0,06 & 3,33 &34 \\ 
\bf 17 &731.294 &76.254  &14,66 &15,03 & 15,29 &9,52 &9,50 & 0,37 &0,26 & 5,17 &0,02 &5,15 &35 \\ 
\bf 18 &787.308 &76.627  & 15,60 &16,28 &- &10,32 &10,78 &0,68 &- &4,82 & -0,47 &5,29 & 38 \\ 
\bf 19 &236.325 &99.667  &15,07 &15,66 &15,81 & 9,71 &10,09 &0,59 &0,15 & 4,98 &-0,38 &5,36 &7  \\ 
\bf 20 &1.107.150 &86.277 &14,77 &15,18 &15,63 &9,99 &9,61 &0,42 &0,44 &5,16 & 0,38 & 4,78 &39 \\ 
\bf 21 &1.206.232 &89.178 &14,26 &14,65 &14,97 &9,44 &9,44 &0,39 &0,33 &4,81 &-0,01 &4,82 &40 \\ 
\bf 22 &178.224 &90.527  & 15,10 &15,66 &15,89 &10,01 &9,77 &0,57 &0,23 &5,33 &0,25 &5,09 &41 \\ 
\bf 23 &522.746 &90.346  &13,97 &14,21 &  14,26 & 10,14 & 9,61 & 0,24 & 0,04 & 4,36 & 0,53 & 3,83 & 42 \\ 
\bf 24 &1.267.375 &91.623 &15,10 &15,65 & 16,02 &10,06 & 9,97 & 0,55 & 0,37 & 5,13 & 0,09 & 5,04 & 8 \\ 
\bf 25 &303.930 &95.689  &14,84 &16,18 & - &  10,47 &10,57 & 1,34 & - & 4,27 & -0,10 & 4,37 & 66 \\ 
\bf 26 &113.660 &96.473  & 11,88 &12,05 &11,18 &9,22 & 10,29 & 0,17 & -0,88 &  1,59 & -1,07 & 2,66 &  68 \\ 
\bf 27 &1.133.270 &119.449 &14,36 &14,94 &15,34 &12,34 & 9,95 & 0,58 & 0,40 & 4,41 & 2,39 & 2,02 & 67 \\ 
\bf 28 &1.080.160 &113.067 &11,00 &11,20 &10,21 & 10,67 & 9,76 & 0,20 & -0,99 & 1,25 &0,91 & 0,33 & 71 \\ 
\bf 29 &61.352 &115.210 &13,52 &13,77 &13,89 & 12,45 & 12,17 & 0,25 & 0,11 &  1,35 &0,28 & 1,07 & 73 \\ 
\bf 30 &666.108 &130.519 &14,21 &14,50 &14,54 & 9,34 &10,65 & 0,29 & 0,03 & 3,56 & -1,31 & 4,87 & 74  \\ 
\bf 31 &879.170 &118.659 &15,92 &16,71 &- & 10,47 & 12,26 & 0,80 & - & 3,65 &  -1,79 &5,45 & 9 \\ 
\bf 32 &1.098.754 &138.812 &15,52 &16,36 &- &10,85 &9,39 &0,84 & - &  6,13 &1,46 & 4,67 &54 \\ 
\bf 33 &344.453 &123.957 &14,04 &14,88 &15,11 &12,46 & 10,91 & 0,84 &  0,23 & 3,13 & 1,55 & 1,58 &24 \\ 
\bf 34 &395.002 &133.480 &13,44 &13,71 &13,73 &11,48 & 10,64 &0,27 & 0,02 & 2,80 &  0,84 & 1,97 & 47 \\ 
\bf 35 &1.209.269 &139.213 &15,71 &16,46 & - &13,02 &10,65 & 0,75 & - &  5,06 &2,37 & 2,69 & 51 \\ 
\bf 36 &469.576 &134.884 &15,65 & 16,41 &- & 13,05 & 9,40 & 0,76 & - &  6,25 & 3,66 & 2,60 & 63 \\ 
\bf 37 &1.166.912 &138.078 &13,27 & 13,56 &  13,49 &12,09 & 12,35 &0,29 &-0,07 & 0,93 & -0,26 &1,19 &50 \\ 
\bf 38 &1.098.582 &141.666 &14,20 & 15,75 & - &10,16 & 13,25 & 1,55 &- &0,95 & -3,09 &4,04 &82 \\ 
\bf 39 &1.293.951 &142.616 &15,31 & 16,02 &-& 9,88 & 11,43 & 0,72 &- &3,88 & -1,54 &5,42 &23 \\ 
\bf 40 &553.801 &143.455 & 15,79 & 16,84 & - & 10,89 &12,78 & 1,05 & - & 3,01 & -1,89 &4,90 &16 \\ 
\bf 41 &1.123.061 &149.094 & 14,70 &16,56 & - & 11,47 & 11,86 & 1,86 &  - &2,84 & -0,40 &3,24 &15 \\
\bf 42 &1.208.210 &168.914 &14,55 & 16,29 &- & 12,06 & 10,13 & 1,74 & - &4,42 &  1,93 &2,49 & 33 \\
\bf 43 &85.035 &155.370 &15,49 &16,41 &- & 11,11 & 10,05 & 0,92 & - &5,45 &1,07 & 4,38 & 29 \\ 
\bf 44 &1.048.087 &156.176 & 16,11 &16,82 & - & 10,36 & 11,51 & 0,72 &- &4,59 & -1,16 & 5,75 &30 \\ 
\bf 45 &1.330.867 &158.737 &9,91 &10,14 &10,56 &12,77 &11,22 &0,23 &0,42 & -1,31 &1,55 &-2,86 &32 \\ 
\bf 46 &1.161.256 &160.856 & 15,34 & 16,01 & - & 9,55 &11,17 &0,67 & - &4,17 &-1,62 &5,79 & 12 \\ 
\bf 47 &1.079.063 &178.168 &15,96 &16,92 &- &11,31 & 10,23 & 0,97 &- &5,73 & 1,08 & 4,65 &46 \\ 
\bf 48 &984.566 &166.573 &14,93 &15,49 &- &10,36 &12,53 &0,57 &- &2,40 & -2,16 &4,56 & 17 \\ 
\bf 49 &529.344 &171.953 &14,34 &16,69 &- &14,65 &10,39 &2,35 &- &3,95 &4,26 &-0,31 &  31 \\ 
\bf 50 &387.820 &178.409 &15,46 &16,24 &- &10,64 &9,96 & 0,77 &- &5,50 &0,68 &4,83 &20 \\ 
\bf 51 &298.220 &180.338 &12,05 &12,24 &11,27 &11,23 &11,24 & 0,18 &-0,97 &0,81 & -0,02 &0,83 & 95 \\ 
\bf 52 &1.160.387 &164.000 &15,02 &15,96 &- &11,60 &10,62 &0,93 &- &4,41 & 0,98 &3,43 &92 \\ 
\bf 53 &1.268.926 &186.612 &14,16 &16,35 &- &15,70 & 11,07 &2,19 &- &3,09 &4,63 &-1,54 &96 \\ 
\bf 54 &425.975 &186.734 &14,91 &16,10 &- &12,09 &11,61 & 1,19 &- & 3,30 &0,48 &2,82 &26 \\ 
\bf 55 &1.193.301 &176.376 &14,89 & 16,00 &15,67 &11,37 &12,08 &1,11 & -0,33 &2,81 &-0,72 &3,53 & 106 \\ 
\bf 56 &940.039 &188.075 &14,57 &15,55 &15,67 &12,20 &13,25 &0,98 & 0,12 &1,32 &-1,05 & 2,37 & 107 \\ 
\bf 57 &102.609 &188.689 &11,65 &11,74 &11,95 &11,39 &15,44 &0,09 &0,21 &-3,79 &-4,06 &0,26 &108 \\ 
\bf 58 &1.111.513 &192.860 & 13,92 &14,24 &14,50 &11,37 & 11,25 & 0,32 &0,25 &2,67 &0,13 &2,55 &97 \\ 
\bf 59 &1.314.305 &201.270 &14,39 & 15,29 &15,43 &11,14 &11,35 &0,90 &0,14 &3,04 &-0,21 & 3,25 & 98 \\ 
\bf 60 &34.901 &203.892 &13,93 &14,23 &14,19 &12,60 &11,71 &0,30 &-0,04 &2,22 &0,89 &1,33 & 99 \\ 
\bf 61 &1.134.698 &204.045 &14,06 & 14,31 &- &11,52 &12,37 &0,25 &- &1,69 &-0,85 &2,54 & 48 \\ 
\bf 62 &1.069.441 &206.801 &12,02 &13,46 &14,83 &9,73 &11,19 &1,44 &1,37 &0,84 &-1,46 & 2,29 & 100 \\ 
\bf 63 &719.612 &207.703 &14,27 & 14,56 &14,49 &13,67 &9,83 &0,29 &-0,07 &4,44 &3,83 & 0,61 &84  \\ 
\hline
\end{tabular}
\end{table}

\begin{table}
\caption{\mbox{Table1: Data on CCD photometry in NGC7788}}
\tiny
\begin{tabular}{|r|r|r|c|c|c|c|c|c|c|c|c|c|r|}
\hline 
\bf ID& \bf X Center& \bf Y Center& \bf Mv & \bf Mb & \bf Mu & \bf Mr & \bf Mi & \bf B-V & \bf U-B & \bf V-I & \bf R-I & \bf V-R &\bf Ref \\
\hline 
\bf 64 &1.008.804 &210.874 & 14,00 & 14,47 &15,11 &9,97 &11,31 &0,47 &0,64 & 2,68 & -1,35 &4,03 &83 \\ 
\bf 65 &549.628 &214.526 &13,94 &13,98 &13,83 &11,04 &9,01 &0,05 & -0,15 &4,93 &2,03 &2,89 &86 \\ 
\bf 66 &952.593 &216.383 & 14,53 &14,93 &15,01 &10,00 &13,36 &0,40 &0,07 & 1,17 &-3,37 &4,53 &81 \\ 
\bf 67 &1.235.482 &220.272 &14,64 &15,20 &15,64 &10,77 &10,02 &0,57 &0,44 &4,62 &0,74 &3,87 &85 \\ 
\bf 68 &474.164 &222.275 &14,94 &15,40 &15,37 &13,83 &10,79 &0,46 &-0,03 &4,14 &3,04 &1,11 &94 \\ 
\bf 69 &517.797 &221.818 &14,72 &16,75 &- &10,36 &9,24 &2,03 &- &5,48 &1,11 &4,37 &60 \\
\bf 70 &1.168.648 &235.176 &14,23 &14,53 &14,52 &10,00 &9,97 &0,30 &-0,01 &4,25 &0,03 &4,22 &62 \\ 
\bf 71 &1.084.598 &209.000 &14,56 &15,02 &14,71 &9,84 &11,39 &0,46 &-0,31 &3,17 &-1,54 &4,71 &69 \\ 
\bf 72 &154.266 &228.003 & 14,41 &17,22 &- & 14,29 &9,85 & 2,80 & - & 4,57 & 4,44 & 0,12 & 70 \\ 
\bf 73 &770.067 &236.289 &  11,35 &11,54 & 10,61 & 10,91 &13,54 & 0,18 & -0,93 & -2,19 & -2,63 & 0,44 & 78 \\ 
\bf 74 &1.188.000 &239.398 &12,75 &13,13 &13,08 &10,82 &9,29 &0,38 & -0,04 & 3,46 & 1,54 & 1,92 &  79  \\ 
\bf 75 &231.610 &236.663 & 14,60 &14,90 & 14,91 & 12,41 &9,97 & 0,30 &  0,01 &4,62 & 2,43 &  2,19 &80 \\ 
\bf 76 &300.440 &242.193 &14,74 &16,61 &- &12,90 &10,75 & 1,87 & - & 3,99 &2,15 & 1,85 & 75 \\ 
\bf 77 &1.164.827 &235.087 &15,55 & 17,13 & - & 9,81 &  9,82 & 1,58 &- & 5,73 & -0,01 & 5,74 & 76 \\ 
\bf 78 &733.342 &247.129 &15,44 &16,44 & - & 10,98 & 13,97 & 1,00 & - &  1,47 &  -3,00 &  4,46 & 77 \\ 
\bf 79 &1.259.108 &249.621 &12,51 & 12,72 & 12,74 &  12,12 &  10,80 & 0,21 & 0,02 & 1,72 & 1,33 & 0,39 & 101 \\ 
\bf 80 &684.657 &255.076 &9,10 & 9,07 & - & 10,89 & 11,92 & -0,02 & - & -2,82 & -1,03 &-1,79 & 102 \\ 
\bf 81 &166.633 &255.921 &14,36 &15,24 &15,29 & 16,11 & 12,68 & 0,88 &0,04 & 1,69 &3,43 &-1,74 & 104 \\ 
\bf 82 &1.083.860 &258.418 &14,26 & 15,42 &15,55 &9,89 & 9,75 & 1,17 &0,13 &4,51 &0,15 &4,37 & 103 \\ 
\bf 83 &581.433 &278.344 &14,45 & 14,93 & 15,25 &10,20 & 10,75 & 0,48 &0,32 &3,70 & -0,55 &4,25 &112 \\ 
\bf 84 &1.141.996 &265.503 &14,31 &15,14 &15,07 &11,69 & 11,96 & 0,83 & -0,07 & 2,35 &-0,27 & 2,62 & 111 \\ 
\bf 85 &1.012.112 &267.119 & 14,92 & 15,38 & 15,45 & 10,85 & 10,75 &0,46 & 0,08 &4,17 & 0,10 & 4,06 & 28 \\ 
\bf 86 &184.264   &276.862 &15,18 &15,78 &15,97 &10,39 &16,56 &0,60 & 0,18 &-1,37 &-6,16 &4,79 &21 \\ 
\bf 87 &719.601   &277.909 & 14,61 &17,33 &- &11,33 & 10,54 & 2,72 & - &4,07 & 0,79 & 3,28 & 105 \\ 
\bf 88 &503.531   &281.766 &14,75 & 15,86 &15,90 & 9,30 &10,11 & 1,11 & 0,04 & 4,64 &  -0,81 & 5,45 &  37 \\ 
\bf 89 &1.280.525 &285.694 &14,83 & 15,21 &  15,13 &  11,00 &  9,91 &  0,38 &  -0,08 & 4,91 & 1,09 & 3,82 &  90 \\ 
\bf 90 &56.364 &286.702 &12,79 & 13,66 &14,43 &11,74 &12,66 &  0,87 & 0,77 & 0,13 &-0,92 &1,05 &113 \\ 
\bf 91 &1.138.311 &287.912 & 10,00 &10,47 &10,37 &10,65 &10,72 &0,47 &-0,10 &-0,72 &-0,07 &-0,65 &109 \\ 
\bf 92 &1.230.722 &288.403 &14,08 &14,32 & 14,49 &10,38 &10,46 &0,24 &0,17 &3,62 &-0,08 &3,70 &110 \\ 
\bf 93 &580.595 &279.772 &16,00 &16,88 &- &14,81 &9,52 &0,88 &- &6,49 &5,29 &1,20 &57  \\ 
\bf 94 &928.899 &295.811 &15,60 &17,07 &- &9,76 &11,65 &1,48 &- &3,95 &-1,89 &5,84 &22  \\ 
\bf 95 &180.055 &304.460 &15,21 &16,51 &- &10,40 &11,05 &1,30 &- &4,15 &-0,65 &4,80 &25  \\ 
\bf 96 &519.837 &319.973 &15,97 &16,91 &- &10,17 &11,43 &0,93 &- &4,54 &-1,26 &5,80 &49  \\ 
\bf 97 &1.198.762 &314.562 &15,47 &16,07 &- &10,74 &10,55 &0,60 &- &4,92 &0,19 &4,74 &52 \\ 
\bf 98 &1.004.401 &317.704 &15,12 &15,65 &15,94 &11,53 &10,42 &0,53 &0,29 &4,70 &1,11 &3,59 &45 \\ 
\bf 99 &1.249.450 &329.871 &15,78 &16,89 &- &9,91 &14,50 &1,11 &- &1,28 &-4,59 &5,87 &65   \\ 
\bf 100 & 1.032.196 &330.772 &15,71 &16,86 &- &10,22 &10,19 &1,15 &- &5,52 &0,03 &5,49 &72   \\ 
\bf 101 &110.870 &336.770 &15,07 &15,60 &15,93 &11,00 &10,30 &0,53 &0,33 &4,78 &0,70 &4,07 &88  \\ 
\bf 102 &629.377 &337.071 &15,15 &15,76 &16,34 &11,06 &10,21 &0,61 &0,57 &4,94 &0,84 &4,10 &61  \\ 
\bf 103 &840.663 &338.015 &14,93 &15,56 &15,86 &11,21 &11,86 &0,63 &0,30 &3,07 &-0,65 &3,71 &91 \\ 
\bf 104 &568.053 &341.255 &14,93 &16,42 &- &10,57 &10,54 &1,49 &- &4,39 &0,03 &4,36 &87 \\ 
\bf 105 &967.031 &346.896 &15,39 &16,00 &- &10,06 &9,95 &0,61 &- &5,45 &0,12 &5,33 &56 \\ 
\bf 106 &936.514 &348.545 &15,62 &16,25 &- &11,84 &10,29 &0,63 &- &5,32 &1,55 &3,78 &55 \\ 
\bf 107 &377.683 &349.869 &15,86 &16,65 &- & 10,15 &11,45 &0,78 &- &4,42 &-1,29 &5,71 &27   \\ 
\bf 108 & 1.314.339 &351.911 &15,52 &16,72 &- &10,54 &11,21 &1,20 &- &4,32 &-0,67 &4,98 &58 \\ 
\bf 109 &177.577 &352.584 &14,92 &15,40 &15,74 &10,47 &10,62 &0,48 &0,34 &4,30 &-0,15 &4,45 &59  \\ 
\bf 110 & 1.136.841 &356.620 &15,10 &17,07 &- &10,72 &11,10 &1,97 &- &4,00 &-0,38 &4,38 &89  \\ 
\bf 111 &664.130 &360.403 &15,90 &16,85 &- &10,09 &9,92 &0,95 &- &5,98 &0,17 &5,81 &64   \\ 
\bf 112 &317.664 &361.486 &15,39 &16,13 &- &12,90 &10,42 &0,74 &- &4,97 &2,47 &2,49 &53  \\ 
\bf 113 &970.000 &360.655 &14,95 &17,17 &- &11,19 &10,06 &2,23 &- &4,89 &1,14 &3,75 &93  \\
\hline
\end{tabular}
\end{table}
\begin{table}
\scriptsize
\mbox{ Table 3: Comparison of NGC7788 cluster parameters with published data} 

\begin{tabular}{|c|c|c|c|c|}
\hline
\bf $E(B-V)$, mag. & \bf $(m-M)_o$ mag.& \bf Distance in parsec & \bf Log Age & \bf Source \\
\hline
0,28$\pm$0,03 &11,9$\pm$0,24 &2398 &7,97 &This paper \cr
0,28 &11,9 &2410 & &Becker  \cr
0,28 &12,03 & &7,59 &Loktin et al.(2001) \cr
0,28 & &2374 &7,59 &Dias et al. (2002) \cr
0,48 &11,87 & &7,48 &Kharchenko et al. (2005) \\
\hline
\end{tabular}
\end{table}

\begin{figure}
{\includegraphics[width=0.45\textwidth]{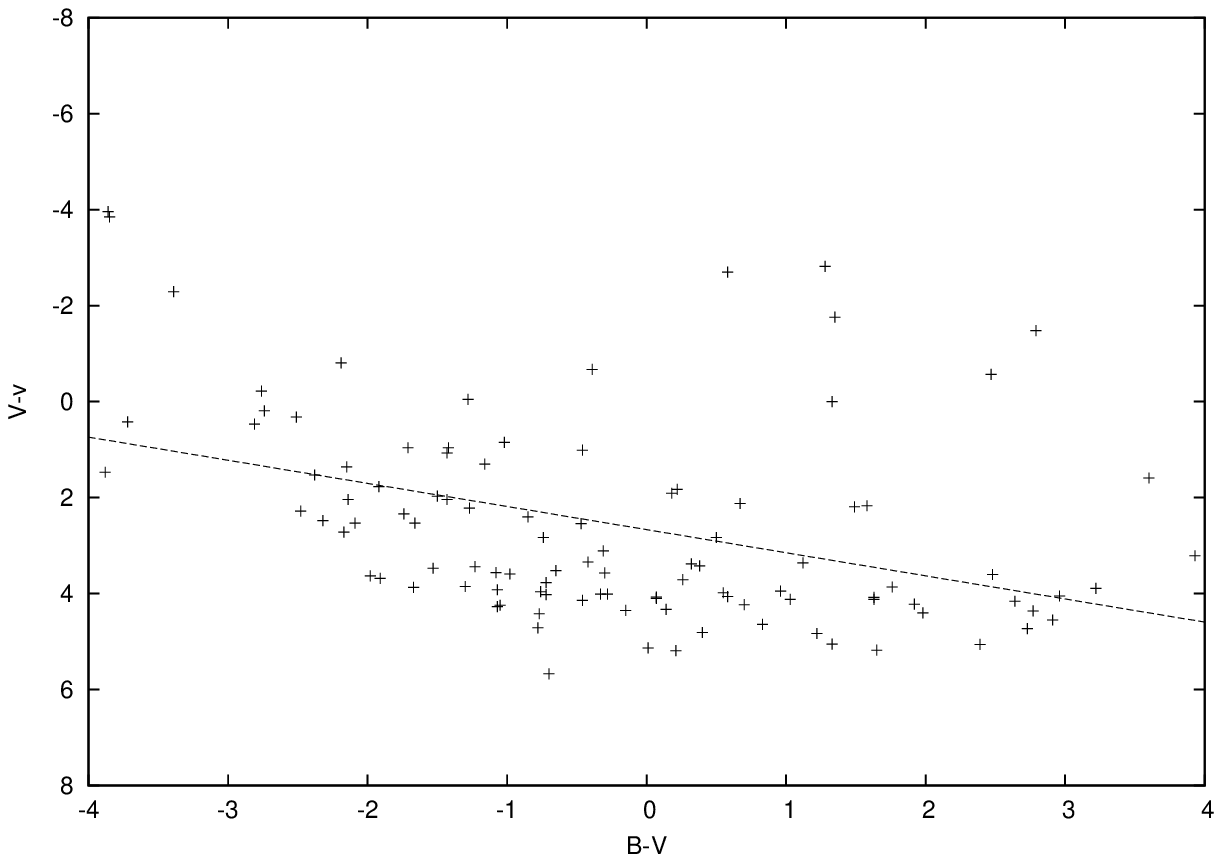}} 
{\includegraphics[width=0.45\textwidth]{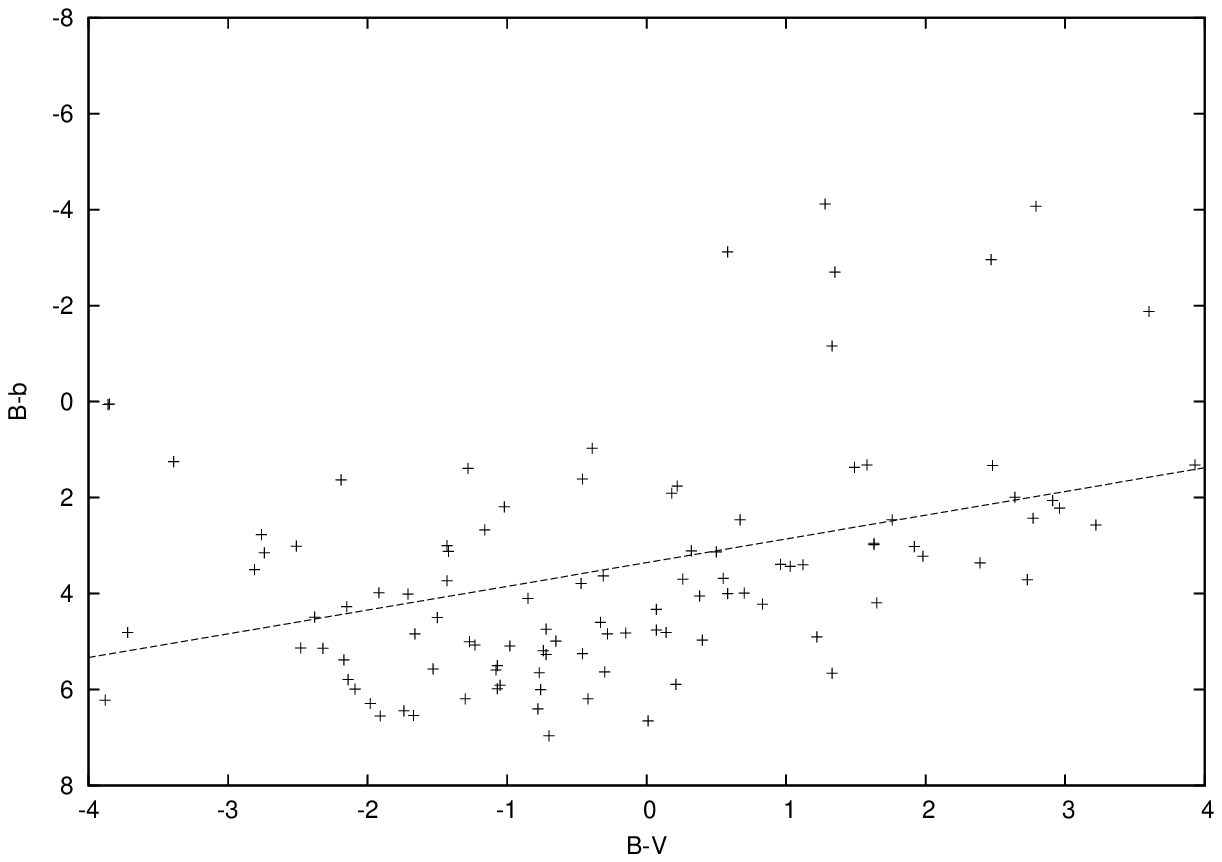}} \\
{\includegraphics[width=0.45\textwidth]{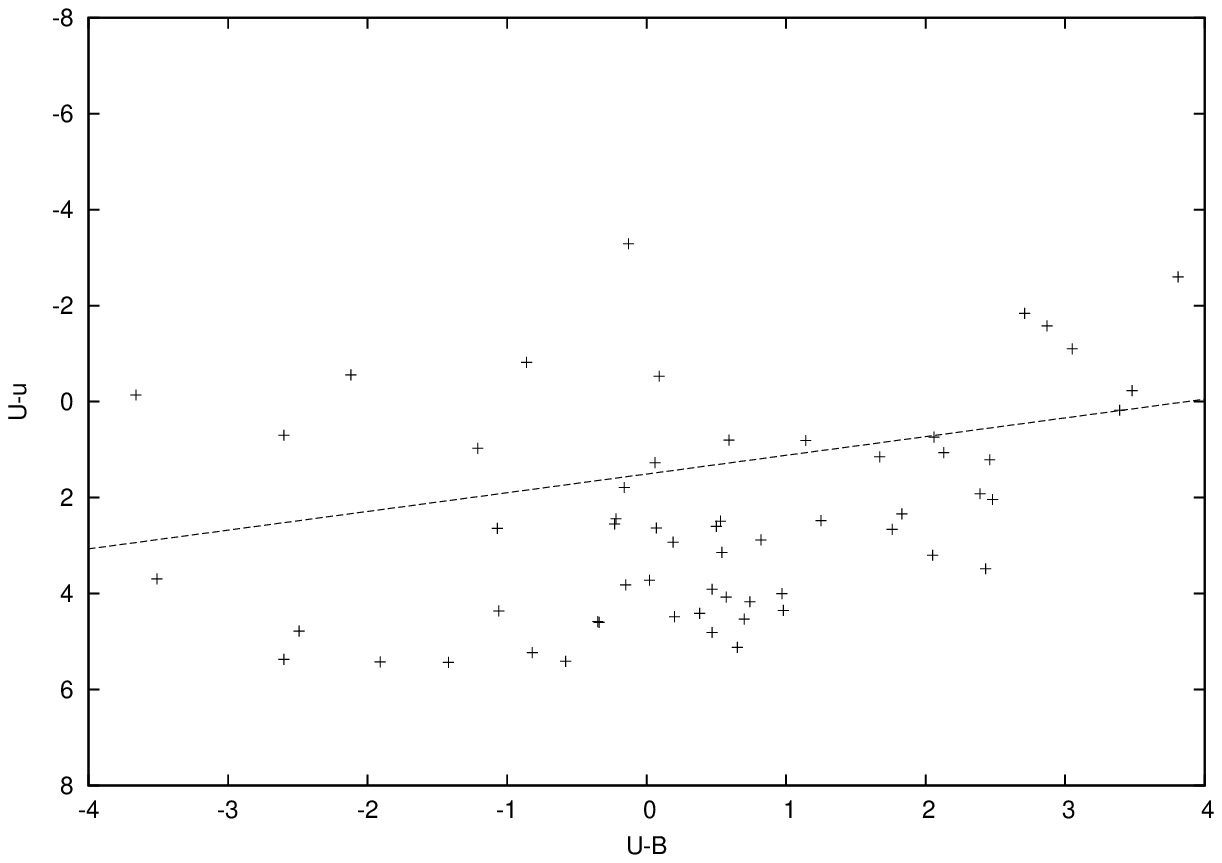}} 
{\includegraphics[width=0.45\textwidth]{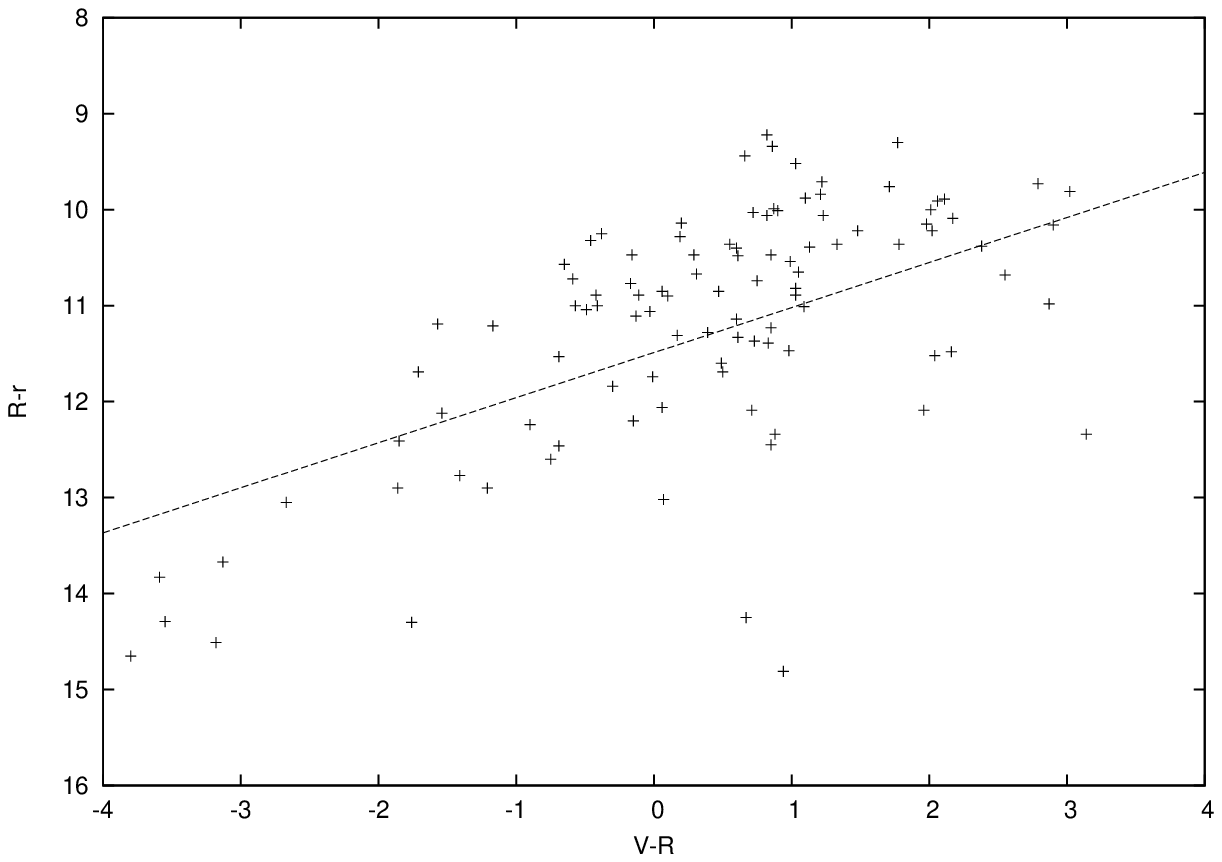}} \\
{\includegraphics[width=0.45\textwidth]{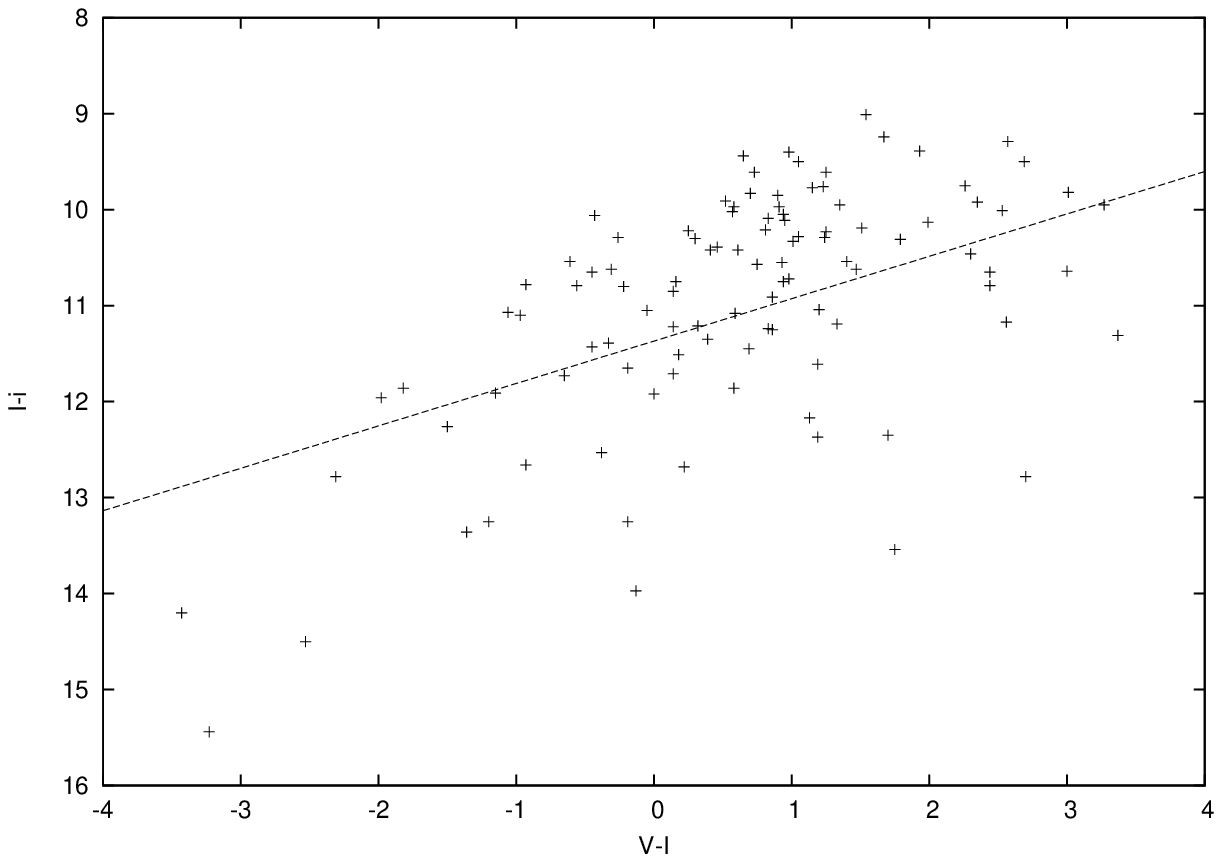}}
\caption{Index of Color Graphics}
\end{figure}

\begin{figure}

{\includegraphics[width=0.45\textwidth]{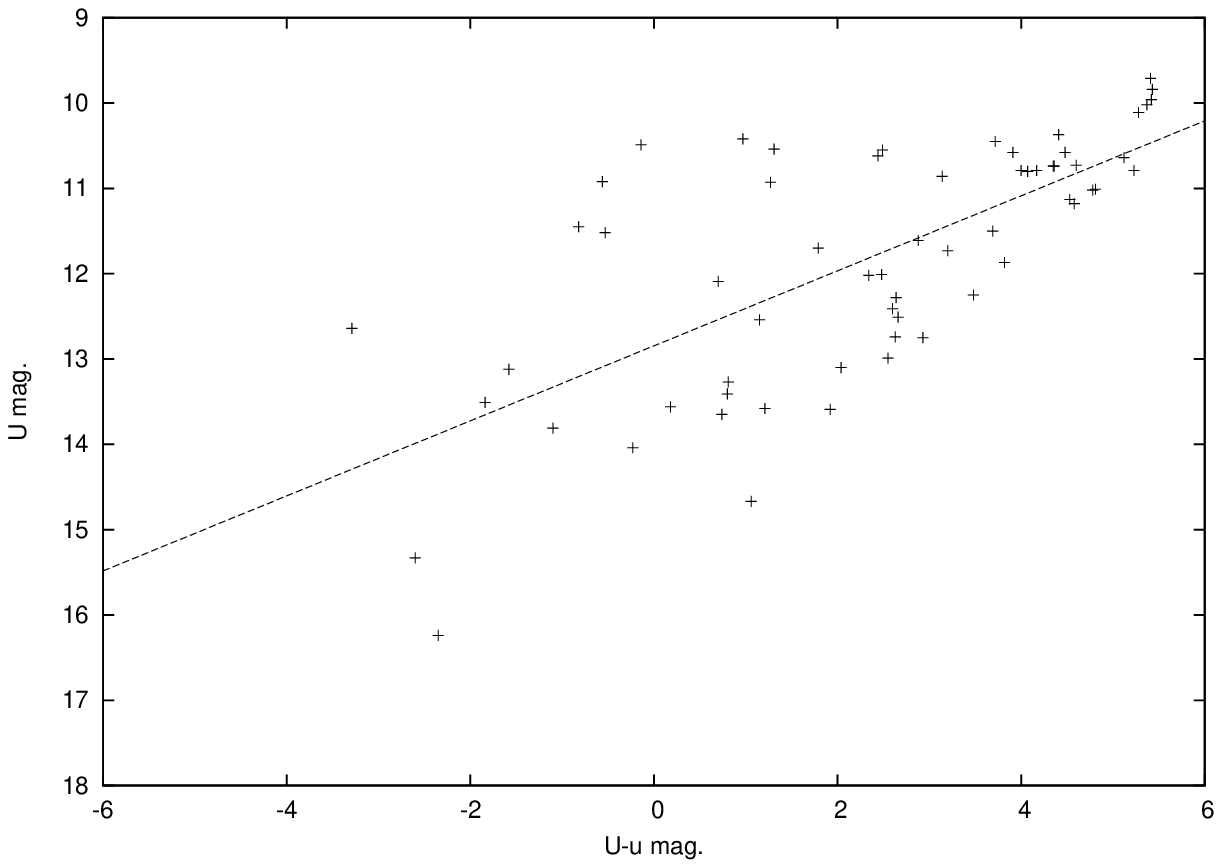}} {\includegraphics[width=0.45\textwidth]{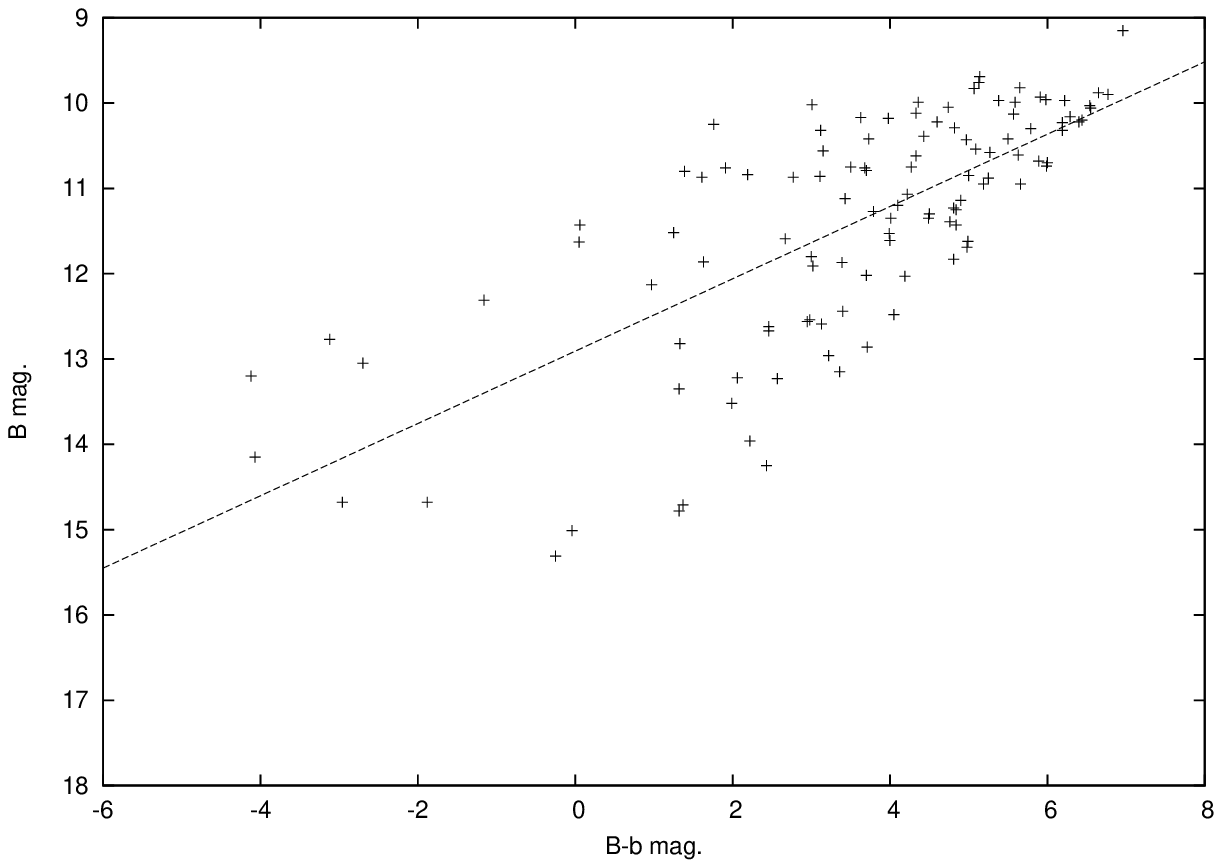}} \\
{\includegraphics[width=0.45\textwidth]{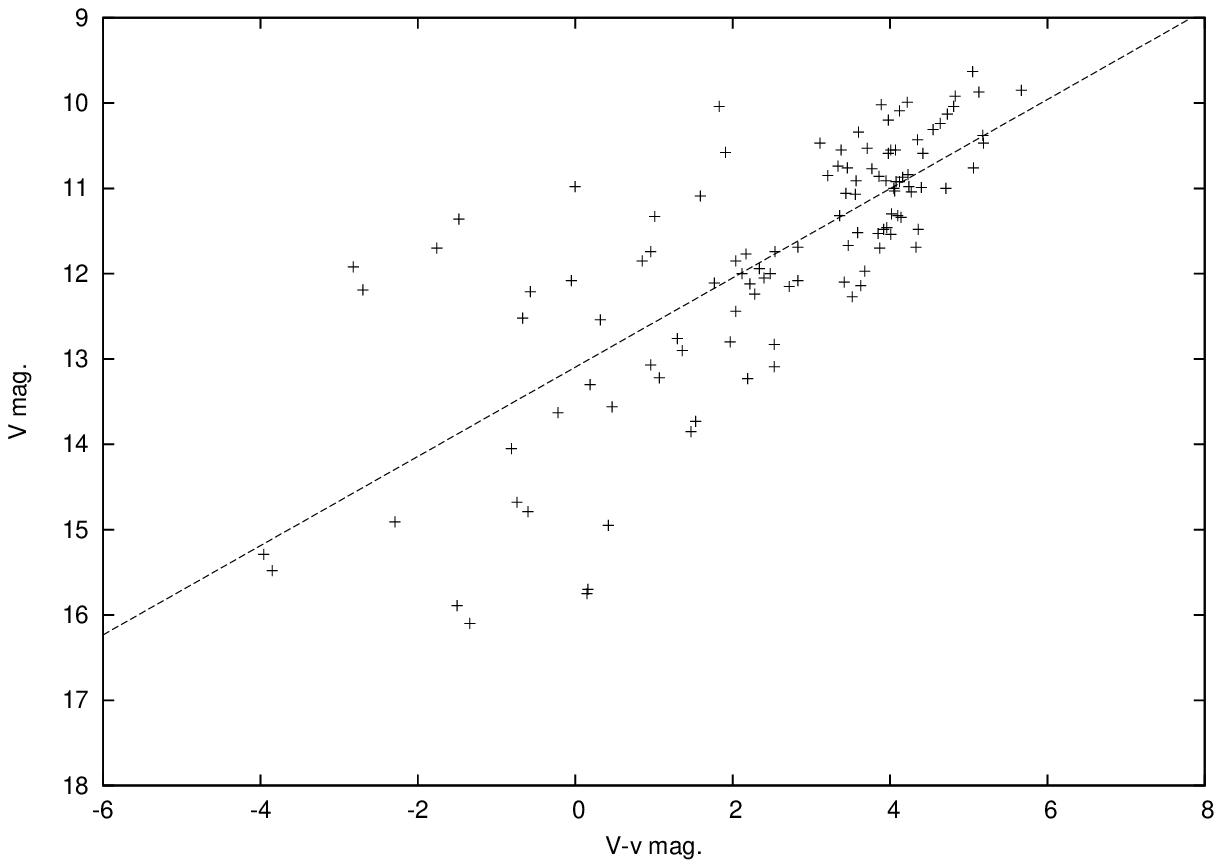}} {\includegraphics[width=0.45\textwidth]{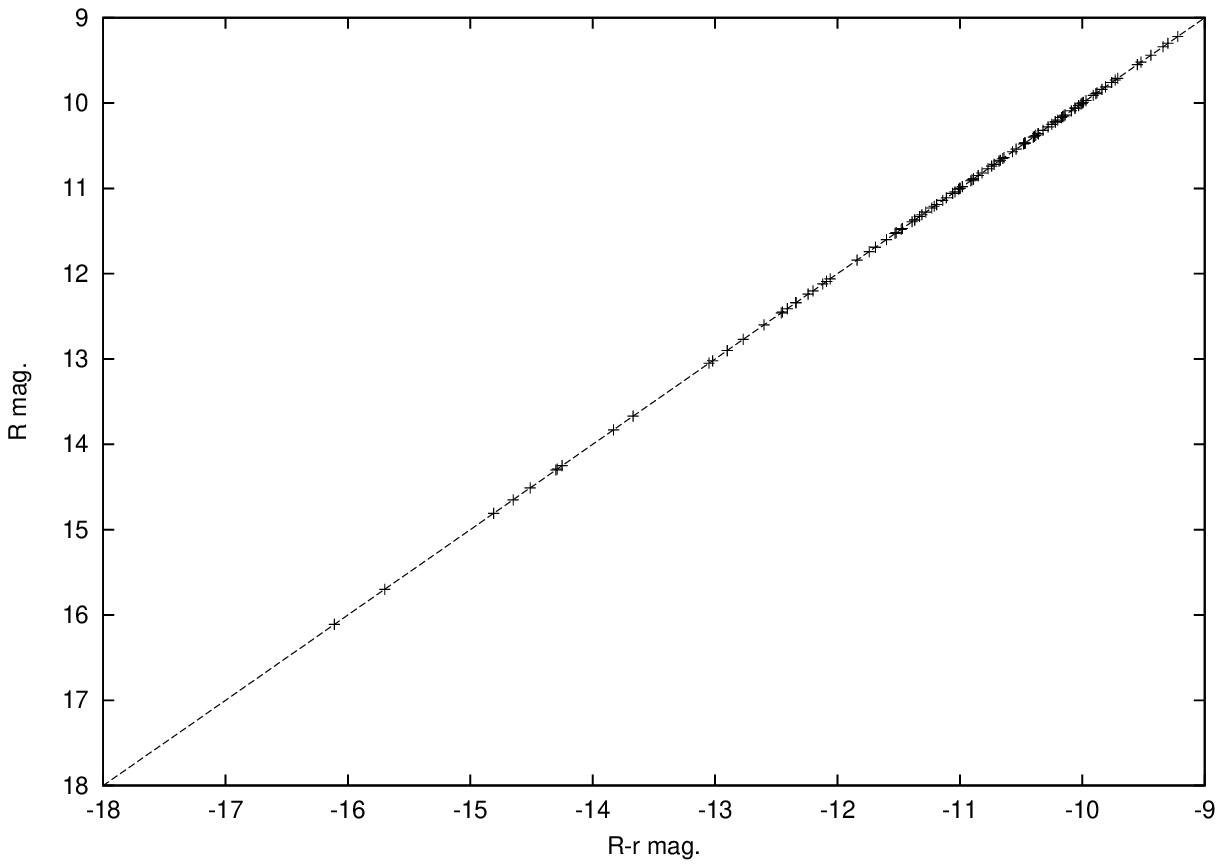}} \\
{\includegraphics[width=0.45\textwidth]{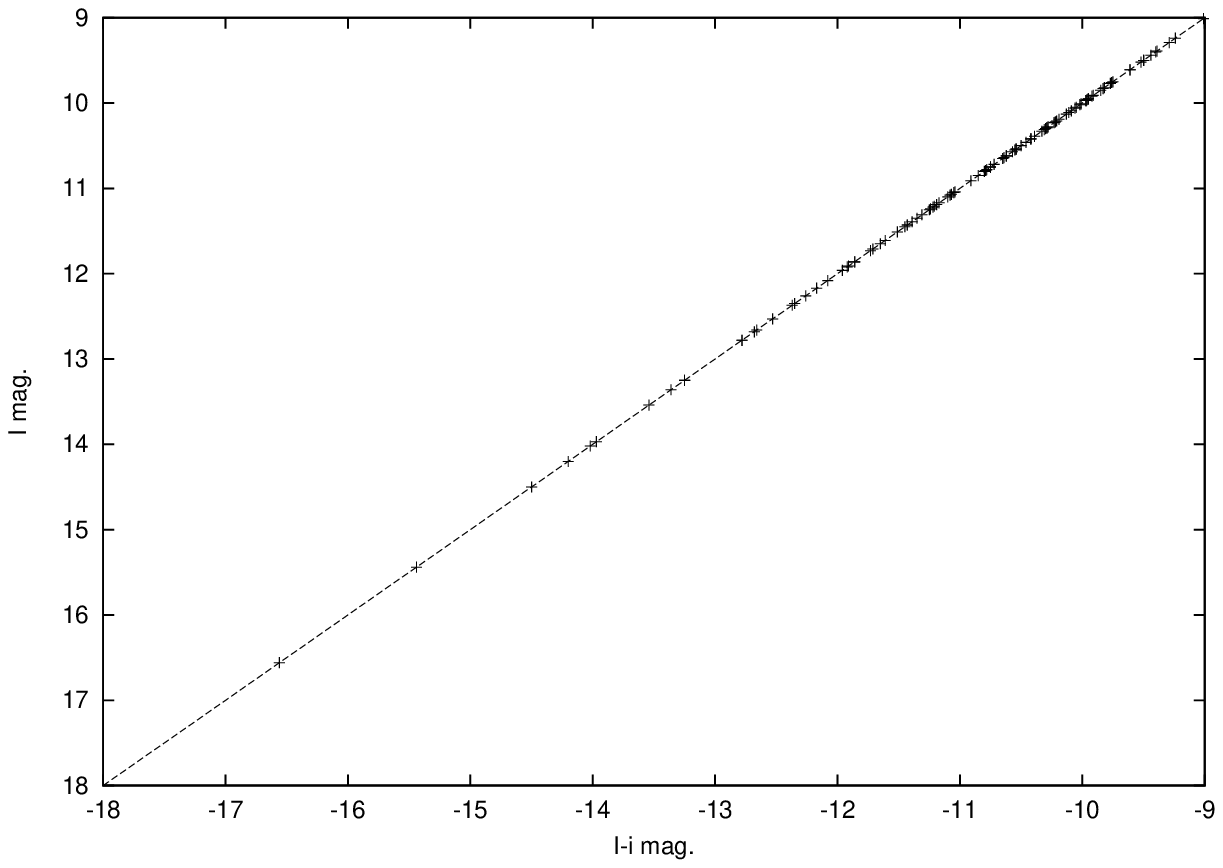}}
\caption{Standard Devition in Magnitude}
\end{figure}

\begin{figure}
\centering
\includegraphics[width=0.90\textwidth]{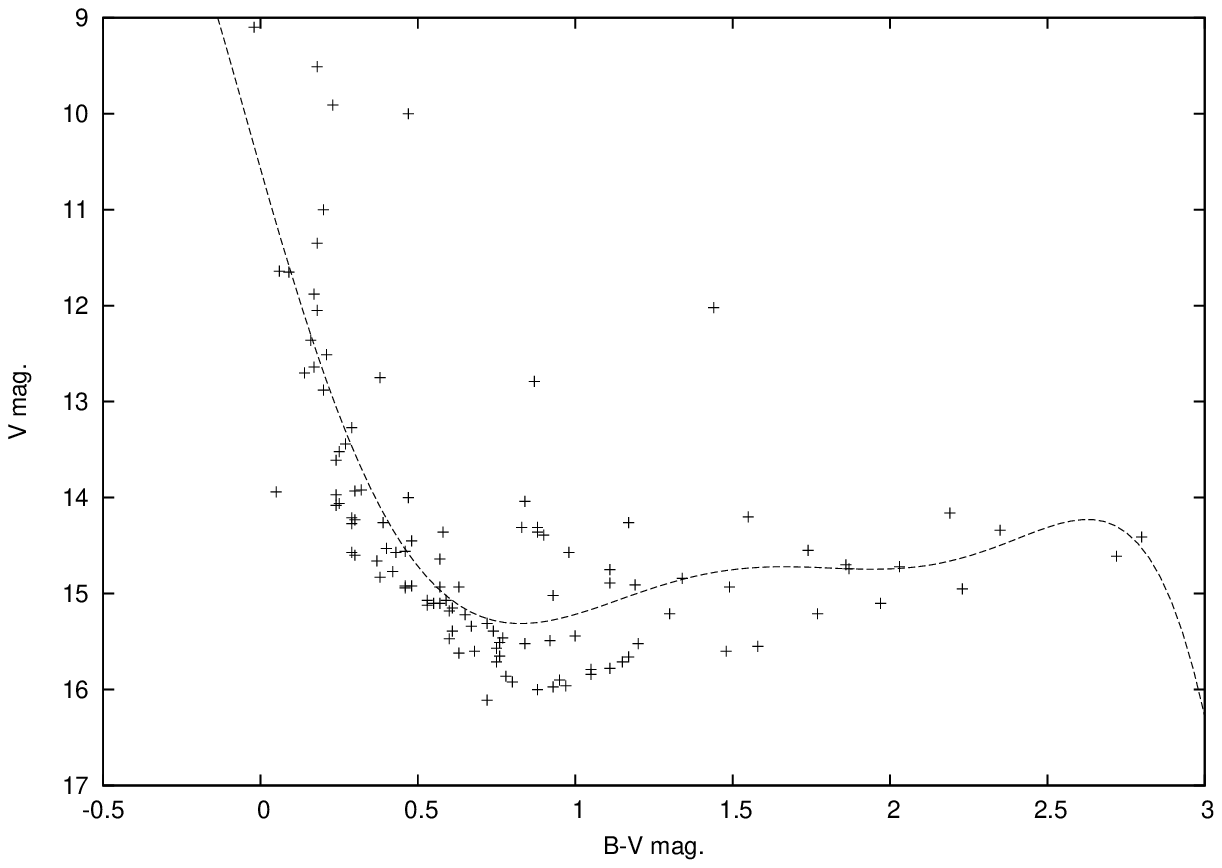} 
\caption{- Colour-Magnitude Diagram (V, B-V) - Black line is Main Sequence}

\end{figure}

\begin{figure}
\centering
\includegraphics[width=0.90\textwidth]{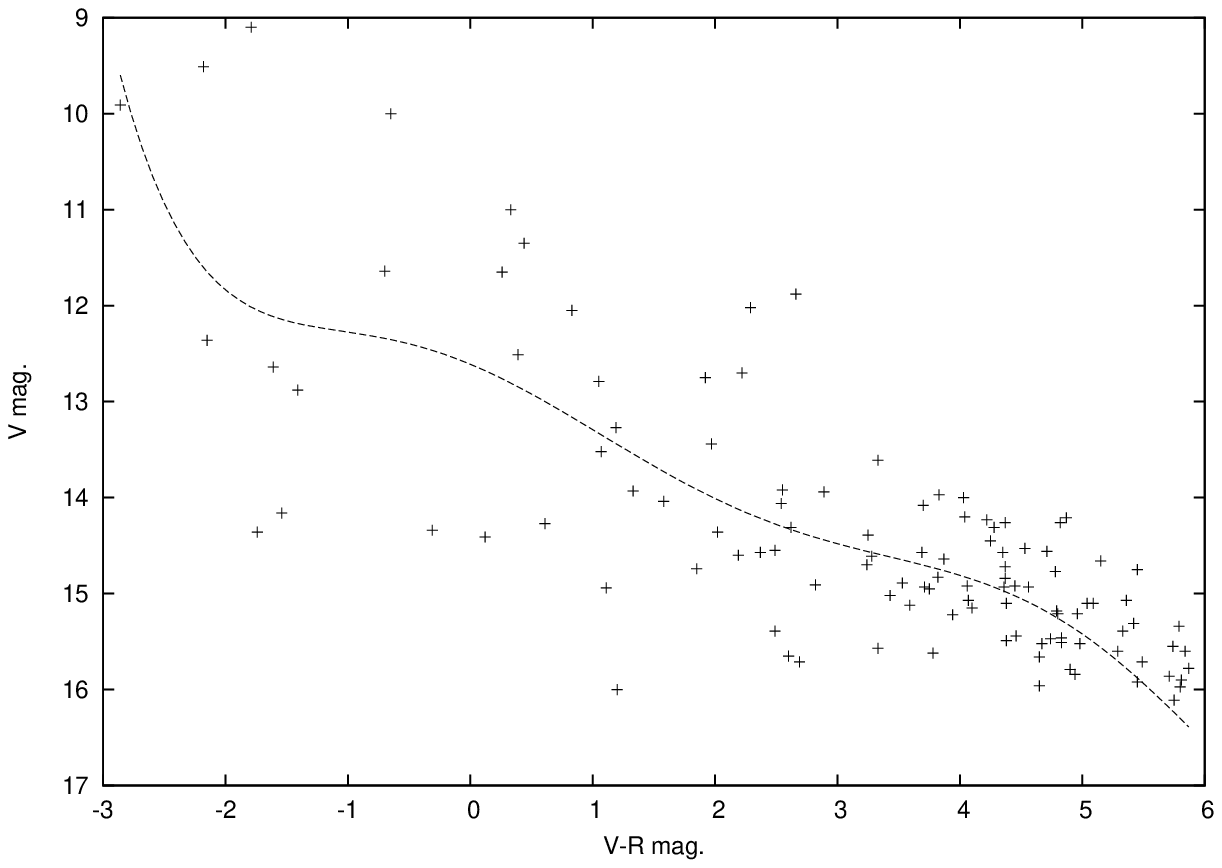} 
\caption{- Colour-Magnitude Diagram (V, V-R) - Black line is Main Sequence}

\end{figure}

\begin{figure}
\centering
\includegraphics[width=0.90\textwidth]{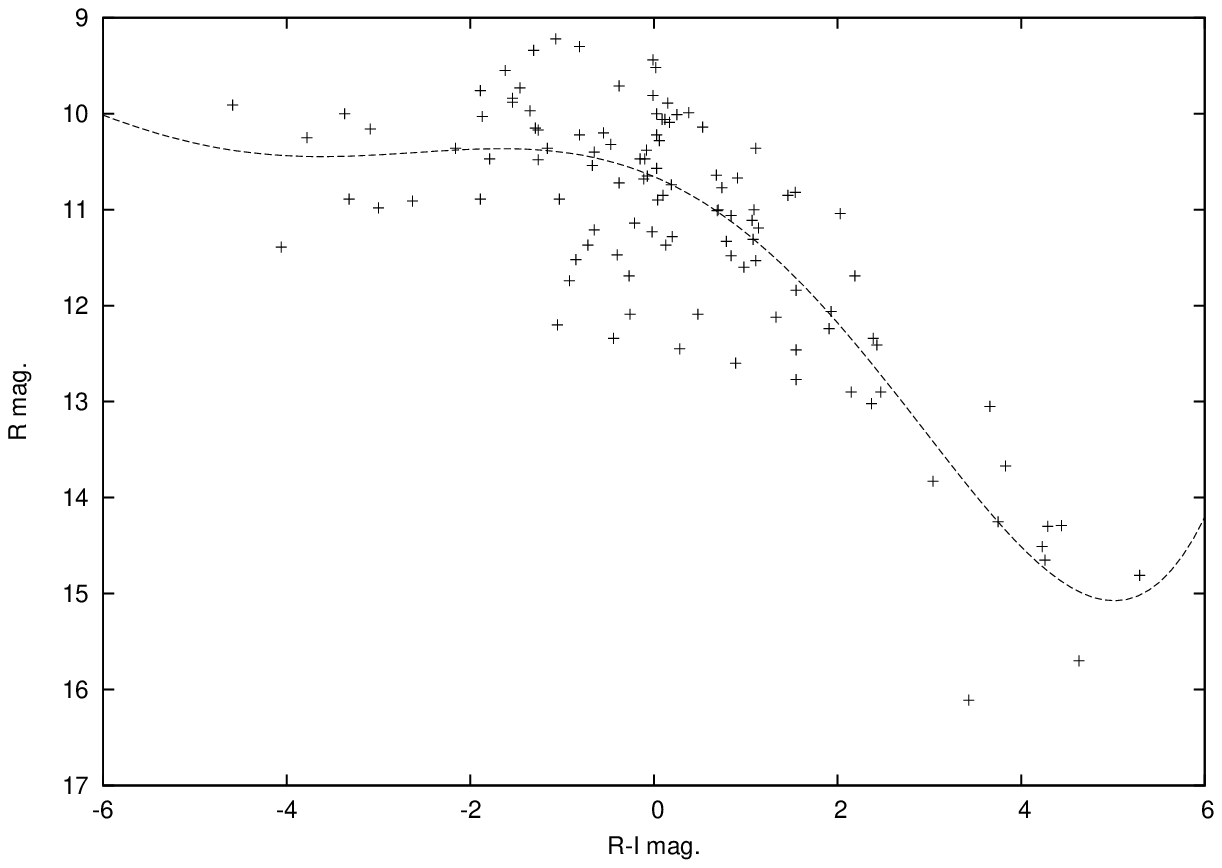} 
\caption{ - Colour-Magnitude Diagram (R, R-I) - Black line is Main Sequence}

\end{figure}

\begin{figure}
\centering
\includegraphics[width=0.90\textwidth]{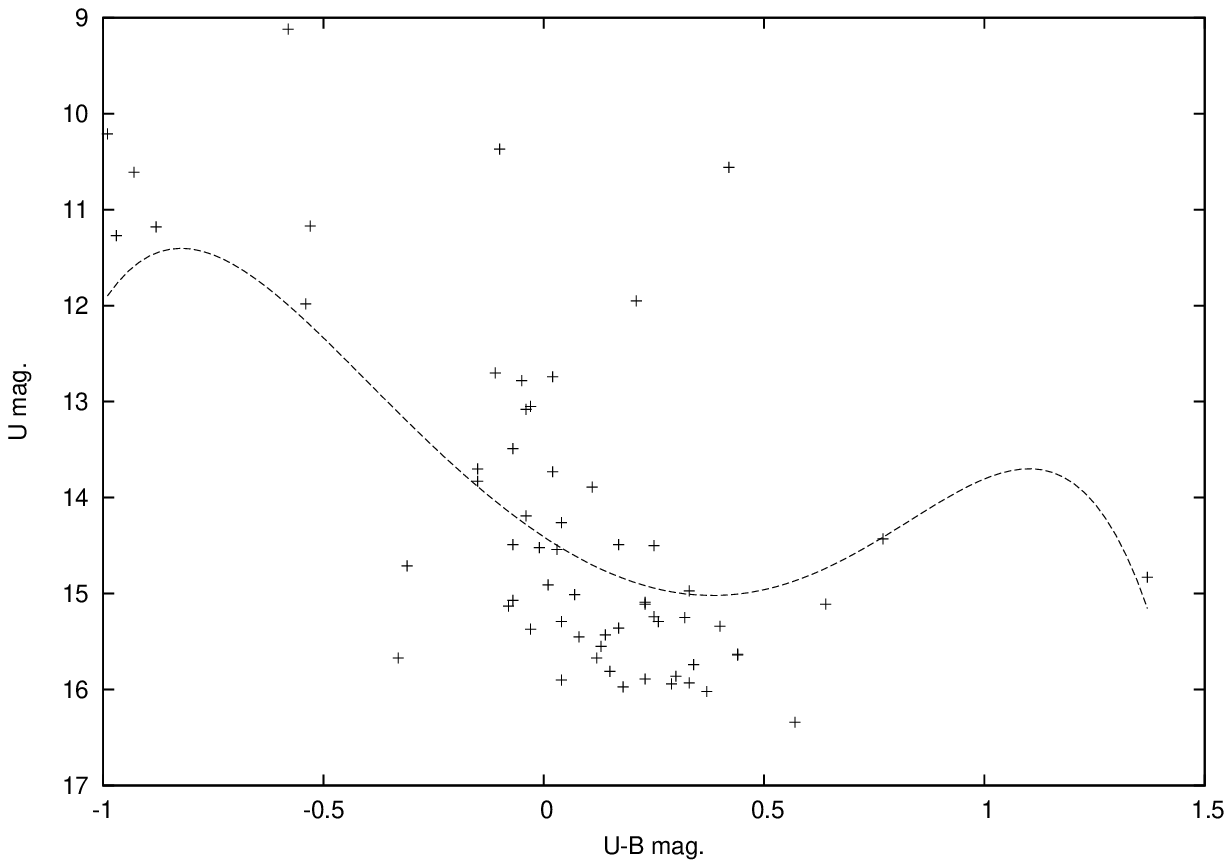} 
\caption{- Colour-Magnitude Diagram (U, U-B) - Black line is Main Sequence} 

\end{figure}

\begin{figure}
\centering
\includegraphics[width=0.90\textwidth]{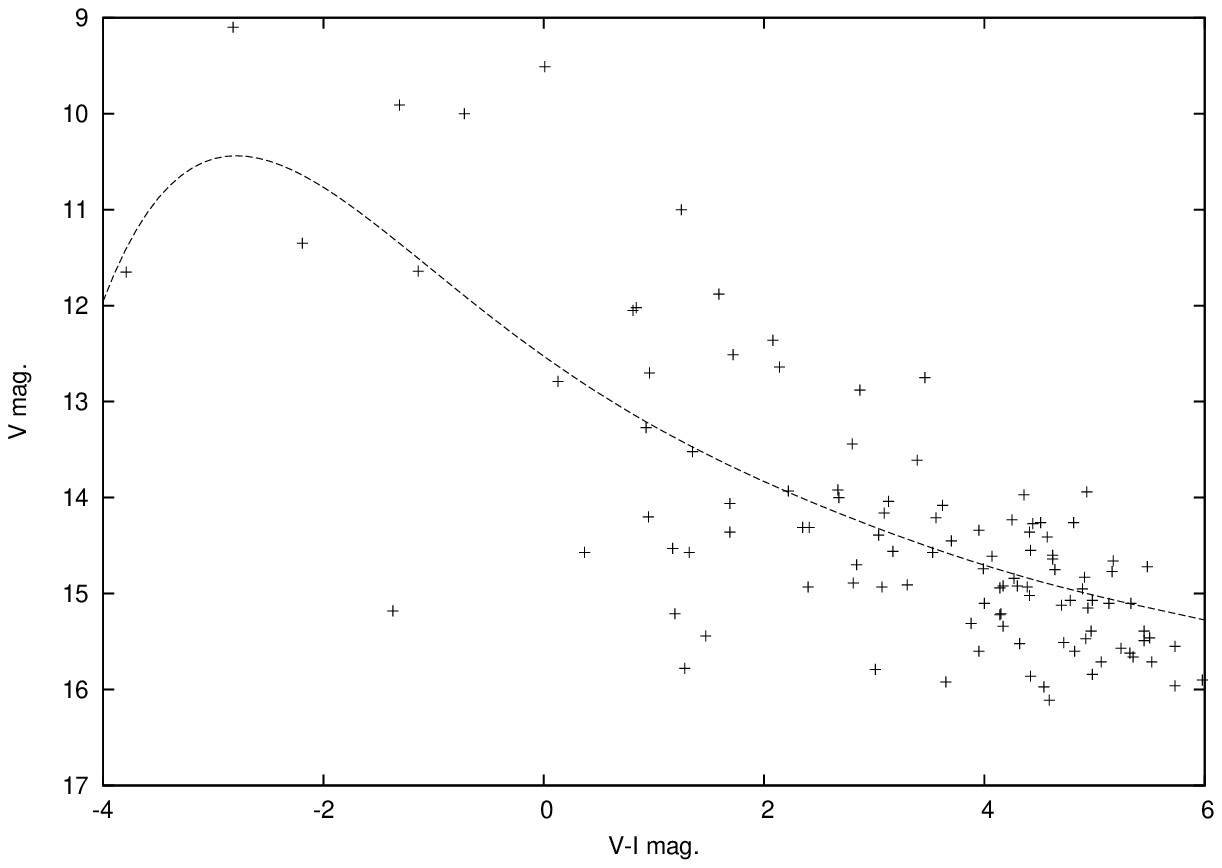} 
\caption{- Colour-Magnitude Diagram (V, V-I) - Black line is Main Sequence}

\end{figure}

\begin{figure}
\centering
\includegraphics[width=0.90\textwidth]{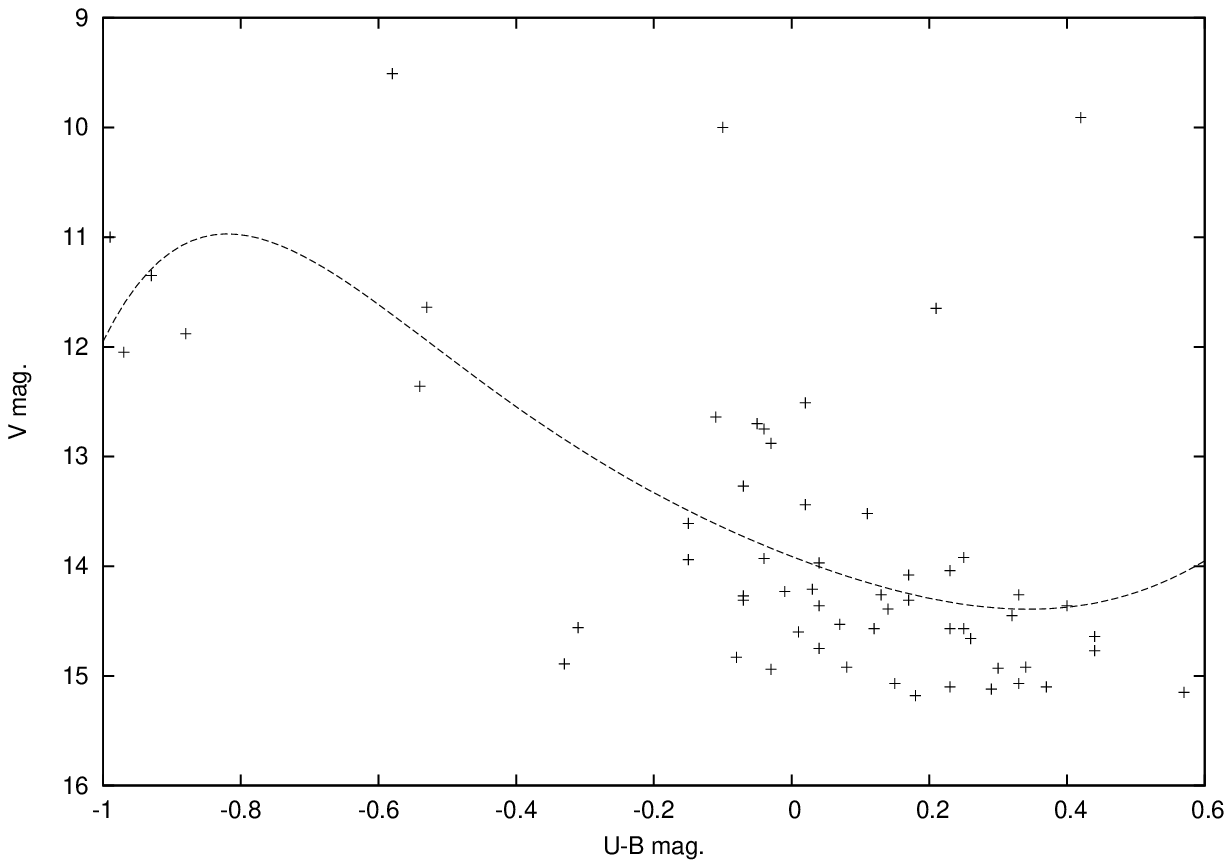} 
\caption{- Colour-Magnitude Diagram (V, U-B) - Black line is Main Sequence}

\end{figure}

\begin{figure}
\centering
\includegraphics[width=0.90\textwidth]{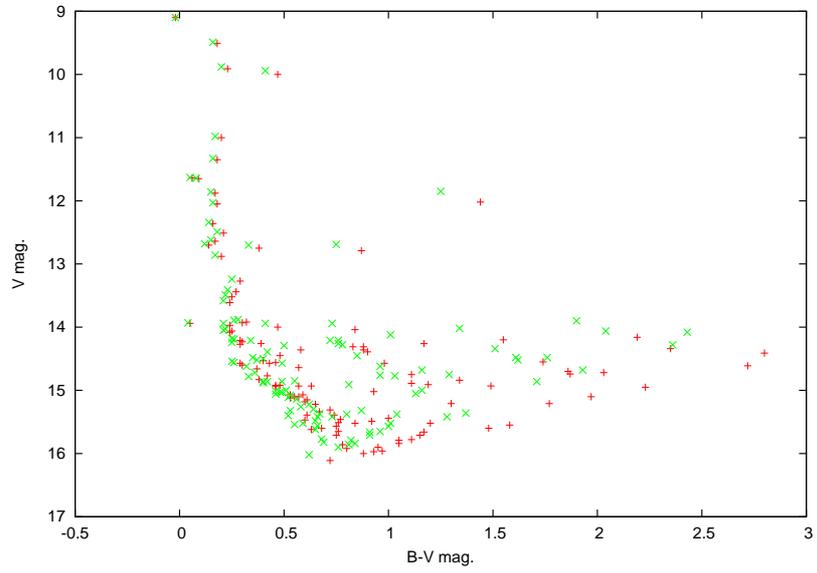} 
\caption{ - Comparison of this study (red) with Becker's (green)}

\end{figure}

\begin{figure}
\centering
\includegraphics[width=0.90\textwidth]{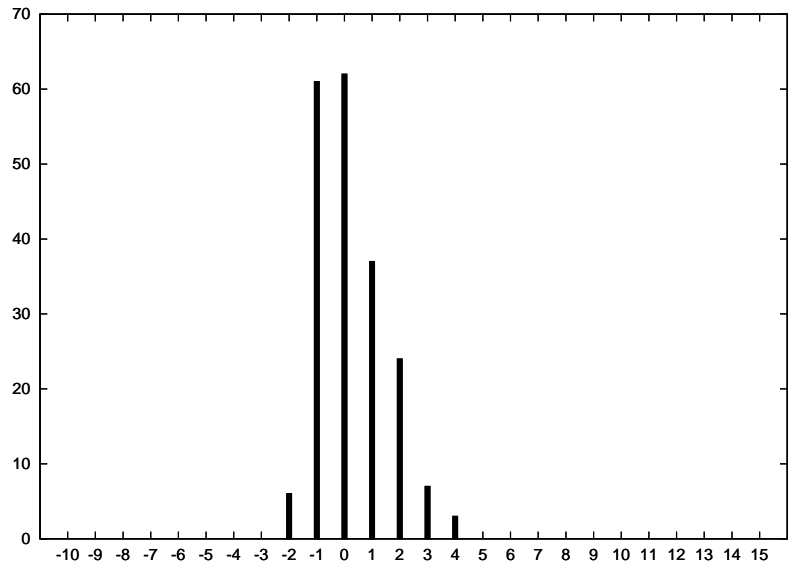} 
\caption{ - The luminosity function of NGC7788 is show as the solid line.}

\end{figure}
\clearpage
\centering

\end{document}